\documentclass{mem}
\usepackage{natbib}
\usepackage{txfonts}
\usepackage{balance}
\usepackage{graphicx}
\usepackage[a4paper,breaklinks,dvipdfm]{hyperref}
\idline{75}{282}

\begin{document}
\def\teff{$T\rm_{eff }$}
\def\kms{$\mathrm {km s}^{-1}$}

\title{Searching for the M+T binary needle in the brown dwarf haystack}

\author{
D. C. \,Bardalez Gagliuffi\inst{1} 
\and A. J. \, Burgasser\inst{1}
\and C. R. \, Gelino\inst{2}
          }


\institute{
Center for Astrophysics and Space Sciences - University of California, San Diego.\\
9500 Gilman Dr., Mail Code 0424, La Jolla, CA 92093, USA.\\
\email{daniella@physics.ucsd.edu}
\and
Division of Physics, Mathematics, and Astronomy - California Institute of Technology.\\
Mail Code 105-24, Pasadena, CA 91102, USA.
}


\authorrunning{Bardalez Gagliuffi}
\titlerunning{M+T binaries}

\abstract{
Multiplicity is a key statistic for understanding the formation of very low mass (VLM) stars and brown dwarfs. Currently, the separation distribution of VLM binaries remains poorly constrained at small separations ($\leq$ 1 AU), leading to uncertainty in the overall binary fraction.  We approach this problem by searching for spectral binaries whose identification is independent of separation.  The combined spectra of these systems exhibit traces of methane imposed on an earlier-type spectra.  When the primary of such a system is a late-type M or early-type L dwarf, however, the relative faintness of the T dwarf secondary (up to 5 magnitudes at K-band) renders these features extremely subtle.  We present a set of spectral indices newly designed to identify these systems, and a spectral fitting method to confirm and characterize them. We apply this method to a library of over 750 spectra from the SpeX Prism Spectral Libraries.  We present new spectral binary candidates, compare them to recent discoveries, and describe ongoing followup to search for resolved companions and/or radial velocity variability.

}
\maketitle{}

\section{Introduction}

The primary mechanisms of brown dwarf formation remain an open question. As opposed to main sequence stars, brown dwarfs require high density regions inside giant molecular clouds to begin their collapse. The main problem is how to stop the accretion before the object grows to a star.  Several mechanisms have been proposed. One theory postulates that brown dwarfs are the result of the turbulent fragmentation of protostellar clouds, where colliding turbulent flows create 2D localized high density regions in which the accretion can be terminated by lack of material in the third direction~\citep{pad02}.  Another theory suggests that brown dwarfs may form by the fragmentation of prestellar disks~\citep{sta09}. If multiple cores form in the same region, the lowest mass member could be ejected by dynamical interaction, preventing the object from gaining mass, and forming a single, isolated brown dwarf~\citep{rei01}.  Finally, the photoerosion of prestellar cores by a nearby OB star through radiation pressure or stellar winds is a case that requires the presence of a massive star~\citep{whi04}.

While all of these theories can successfully form brown dwarfs, it is not clear which one prevails.  One way to examine this is by studying multiplicity. Turbulent fragmentation favors the formation of systems of several brown dwarfs; disk fragmentation tends to form single wide brown dwarf companions to main sequence stars; ejection of prestellar cores and photoerosion prefers the formation of isolated brown dwarfs. About $\sim$ 100 VLM stars have been discovered to date, which correspond to a binary fraction of $\sim$15\%~\citep{bur06}.  While the low fraction would seem to favor the ejection or photoerosion theories, most known systems were identified using high-resolution imaging, and thus are limited by the angular resolution of the instruments.  We can avoid this bias by studying spectral binaries.

Spectral binaries are systems of two spectrally unresolved objects (due to their small angular separation) whose blended-light spectra shows features from both. In this study, we are looking for T brown dwarf features in late M and early L dwarf spectra. Since M dwarfs are the most common objects in the galaxy, and the brightest low-mass bodies, observing them allows us to survey a large volume of the sky. In particular, since the spectra of M and T dwarfs are distinctive, we focus on a methane absorption band at 1.6~$\mu$m, which is typical of T dwarfs, but never found in M dwarfs since their atmospheres are not cool enough to harbor CH$_4$. 


\section{SpeX Spectral Sample}

We started with a sample of 989 low-resolution (R$\sim$100), near-IR spectra of M, L and T dwarfs from the SpeX Prism Spectral Libraries.  The data span wavelengths of 0.8-2.3~$\mu$m, and were obtained over the last 13 years from the SpeX spectrograph, located at the Infrared Telescope Facility in Mauna Kea, Hawaii~\citep{ray03}. This sample excludes known binaries, but includes subdwarfs, unusually blue/red dwarfs, and young brown dwarfs. The entire sample was used for the template fitting, but only 769 M7-L7 dwarfs were used to search for spectral binaries. 


\section{Identifying new binaries}

\begin{figure}[h]
\resizebox{\hsize}{!}{\includegraphics[clip=true]{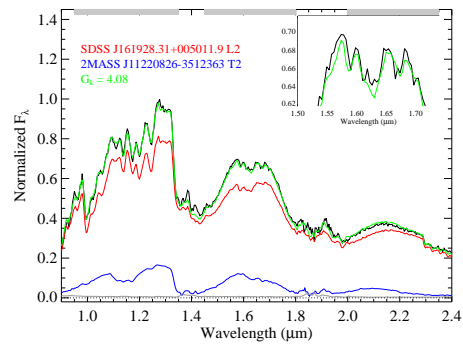}}
\caption{
\footnotesize
This figure shows in black the spectrum of 2MASS J0931+2802, a strong binary candidate with component spectral types of L2 and T2. Early-L and a early-T spectra (red and blue, respectively) are added to create a binary template (green), which proved to be a significantly better fit than any single L dwarf template.}
\label{fig:binaryfit}
\end{figure}

First, we visually selected binary candidates by searching for indications of methane absorption. We specifically looked for a ``dip'' around $1.6~\mu$m, where the left wall is due to flux from the companion and the right side comes from the primary. As can be seen in Figure~\ref{fig:binaryfit}, there is a clear concavity in the H band, indicating the presence of CH$_4$ in an early-L spectrum.

\begin{figure}[h]
\resizebox{\hsize}{!}{\includegraphics[clip=true]{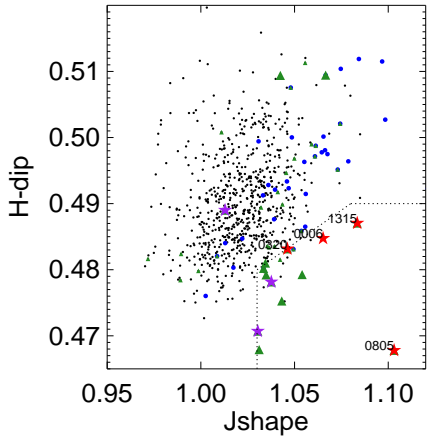}}
\caption{
\footnotesize
One of the 11 parameter spaces showing a combination of spectral indices.  The red stars indicate the benchmarks, the blue circles are unusually blue objects, while the large and small green triangles correspond to strong and weak candidates.}
\label{fig:spexindex}
\end{figure}

Since visual examination can be biased, we also looked for binaries in a more quantitative way using spectral indices. We used indices previously defined by~\citet{bur02} and~\citet{bur10}, customized to find L+T binaries, as well as three new indices, designed for detecting M+T binaries. The new indices were determined by examining the spectra of previously confirmed M+T binaries.  After subtracting single template spectra, we chose the regions showing the greatest contrast and designed three spectral indices accordingly.  Comparing all 11 indices and spectral types, we identified the 11 pairs that best distinguished known binaries, an example of which is shown in Figure~\ref{fig:spexindex}.  After defining regions that segregated our benchmark binaries, we identified 9 strong and 23 weak binary candidates.

In order to verify and characterize the binary candidates, we fit both single and binary templates.  The latter were assembled by constraining the spectral type of the primary to M7-L7 and the secondary to L8-T8, resulting in a total of 65662 binary templates.  The candidate spectra were compared to each single and binary template, and the best fits were ranked by their $\chi^2$.  Finally, an F-test statistic was used to compare the best single and binary fits to test the null hypothesis (that the source is single), and as a weighting factor to determine the average spectral type of each single and binary components.

\section{Results}
\begin{figure}[h]
\resizebox{\hsize}{!}{\includegraphics[clip=true]{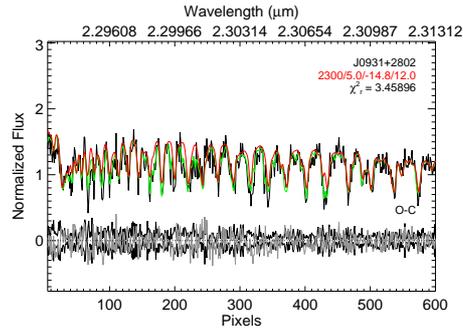}}
\caption{
\footnotesize
Forward-modeling fit of a high-resolution Keck/NIRSPEC spectrum of the binary candidate 2MASS J0931+2802. The center black line is the data; the red line is the best model fit; the green line includes telluric absorption.  The bottom spectrum displays the residuals (black) and the uncertainties (gray).  We extract RV, Vsin$i$ and atmospheric parameters from this fit.}
\label{fig:hires}
\end{figure}

From the original sample of $\sim$1000 low-resolution NIR spectra, we identified 16 visual candidates, and 9 strong and 23 weak candidates from spectral indices. Of these, 7 appear to be promising binary candidates based on spectral template fits.  We are now conducting high resolution imaging and spectroscopic monitoring observations to confirm their binary status and measure orbit properties, as shown in Figure~\ref{fig:hires}.

\begin{acknowledgements}
The authors thank the IRTF staff and observers, particularly John Rayner and Michael Cushing, for their technical support during the observations.
\end{acknowledgements}

\bibliographystyle{aa}

\end{document}